\documentclass{article}
\usepackage[utf8]{inputenc}
\usepackage{amssymb}
\usepackage{amsmath}
\usepackage{amsfonts}
\usepackage{mathtools}
\usepackage{bbm}
\usepackage{xcolor}
\usepackage{tensor}
\usepackage{booktabs}
\usepackage{geometry}
\title{Non-singular Kerr-NUT-de Sitter spacetimes}

\author{Jerzy Lewandowski \thanks{Jerzy.Lewandowski@fuw.edu.pl} \and Maciej Ossowski \thanks{Maciej.Ossowski@fuw.edu.pl}}

\date{%
\small
    Faculty of Physics, University of Warsaw,\\
ul. Pasteura 5, 02-093 Warsaw, Poland\\[2ex]%
    \today
}

\begin{document}

\maketitle

\begin{abstract}
We study Killing horizons and their neighbourhoods in the Kerr–NUT–(anti-)de Sitter and the accelerated Kerr–NUT–(anti-)de Sitter spacetimes. 
The geometries of the horizons have an irremovable singularity at one of the poles, unless the parameters characterising the spacetimes satisfy the constraint we derive and solve in the current paper.
In the Kerr-NUT-de Sitter case, the constraint relates the cosmological constant of spacetime and the horizon area, leaving 3 parameters free.
In the accelerated case the acceleration becomes a 4th parameter that allows the cosmological constant to take arbitrary value, independently of the area.
We find that the neighbourhoods of the non-singular horizons are non-singular too, at least in the non-extremal case.
Finally, we compare the embedded horizons with previously unembedded horizons provided by the local theory of type D Killing horizons to the second order. 
\end{abstract}

\tableofcontents

\section{Introduction}
The Petrov type D spacetimes with the NUT parameter $l$ are known to posses an infinite, string-like singularity located along the rotational symmetry axis \cite{griffiths_podolsky_2009}. 
With an appropriate choice of the coordinates the singularity can be reduced to be only a semi-infinite string contained in half of the axis. 
In the case of the Taub-NUT spacetime Misner proposed a non-singular interpretation, where one suitably glues two non-singular patches of spacetime \cite{misner}.
The price is topological non-triviality - the spacetime becomes topologically $S^3\times \mathbb{R}$. 
In the general case of the Kerr-NUT spacetime, this is not sufficient, and still a conical singularity passing along all the spacetime persists \cite{griffiths_podolsky_2009}.
The properties of the Kerr–NUT–(anti-)de Sitter spacetimes and their relation to the Kerr family has been extensively studied in the literature \cite{Millerdoi:10.1063/1.1666343,Mars2013,Mars2016,Mars2016b,dadhich2002}.
\textcolor{black}{Recently a novel approach to interpret the conical singularity free region of Kerr-NUT-de Sitter spacetimes as a bouncing, non-singular cosmological solution was proposed \cite{Anabalon2019}}.
In the current paper, we address this problem in a new way, by focusing on Killing horizons and their neighbourhoods, and using coordinate independent geometric methods.
Given a Kerr–NUT–(anti-)de Sitter spacetime we choose a Killing horizon therein, determined by a root $r_0$ of a suitable metric component, and the corresponding Killing vector. 
The horizon can be extended to a non-trivial U(1) bundle over $S^2$ that does not admit a global cross-section.
However, the orbit space $S^2$ inherits a unique spacelike geometry from the spacetime.
The geometry has a conical singularity at (at least) one of the poles, for a generic value of the four parameters $(m, a, l, \Lambda)$ defining the Kerr–NUT–(anti-)de Sitter spacetime.
What we find is that the singularity can be removed if and only if the cosmological constant $\Lambda$ is suitably adjusted to the remaining parameters $(m, a, l)$, namely 
\begin{equation} 
\label{eq:constraint-params}
        \frac{3}{\Lambda}=a^2+2l^2+2r_0^2, \quad R^2=\frac{3}{2\Lambda}.
\end{equation}
where $r_0$ can be obtained for a given $m$  and $R^2$ is the area radius.
We also study the geometry of the 3-dimensional orbit space of the Killing vector field in a future/past neighbourhood of the horizon \cite{chrusciel}.
The neighbourhood topology is 
\begin{equation}
    \label{eq:bundle}
    S^3\times \mathbb{R}.
\end{equation}
The orbit space topology $S^2\times \mathbb{R}$ and the geometry inherited from the spacetime are uniquely defined.
We show that the orbit space geometry is smooth at every point for a sufficiently thin neighbourhood in the case if (\ref{eq:constraint-params}) is satisfied.
This result applies also to the spacetime geometry of the neighbourhood. 
  
Next, we generalise our result to the accelerated Kerr-NUT-(Anti) de Sitter spacetimes characterised by five parameters $(m, a, l, \alpha, \Lambda)$.
They are a five dimensional subfamily of the Plebański-Demiański family describing a rotating and accelerating black hole with a NUT parameter on the cosmological constant background.
We find a four dimensional subfamily that contains a singularity free horizon of a singularity free neighbourhood.

The idea of the current research originates from the program of abstractly defined vacuum isolated horizons (IH) of the Petrov type D with a cosmological constant \cite{geometryhorizonsAshtekar_2002,MechanicsIHPhysRevD.64.044016,GeometricCharacterizationsoftheKerrIsolatedHorizon,Lewandowski_2003,Lewandowski_2006,DOBKOWSKIRYLKO2018415,localnohairPhysRevD.98.024008,hopf}. A comprehensive review of the IH theory and its application can be found here \cite{Ashtekar2004}.
They can be characterised with respect to the principal fiber bundle structure of the null Killing flow.
In the trivial bundle case, all the axisymmetric type D IHs set a three dimensional family, and each of them corresponds to a Kerr–(anti-)de Sitter spacetime \cite{localnohairPhysRevD.98.024008} (modulo subtleties that arise in the extremal horizon case).
In the case, when the bundle structure is that of a non-trivial U(1) principal fiber bundle over $S^2$, a four dimensional family of the axisymmetric type D IHs was found \cite{hopf}.
Despite obvious expectations, those type D IHs generically are not embeddable in any of the Kerr–NUT–(anti-)de Sitter spacetimes \cite{hopf}.
This issue is addressed in the last section of the current paper.
We find the explicit correspondence between the non-singular horizons in Kerr-NUT-de Sitter spacetimes ($\Lambda>0$ turns out to be one of necessary conditions) found in the current paper and a three dimensional subfamily of the type D horizons originating from the abstract theory of the type D equation \cite{hopf}.
The non-singular horizons contained in the accelerated Kerr–NUT–(anti-)de Sitter spacetimes also have a topology of a non-trivial bundle over $S^2$ and thus they must also correspond to the type D horizons obtained as the direct solutions to the type D equation.
However, an explicit $1-1$ map between the horizons has not been constructed.

\textbf{Notation.}
If the coordinates are not given explicitly we use the following index conventions: 
\begin{itemize}
    \item Lowercase Greek letters correspond to spacetime tensorial indices: $\alpha,\beta,\gamma\dots=0,1,2,3$
    \item Lowercase Latin letters correspond to tensorial indices on a 3-manifold: $i,j,k\dotsc=1,2,3$.
\end{itemize}
The Einstein summation convention is assumed.

\section{Kerr–NUT–(anti-)de Sitter spacetimes}
In our paper we consider the Kerr-NUT spacetimes with a cosmological constant $\Lambda$ and in Sec. \ref{sec:gen-bh} its accelerated generalisation.
In the current short section we recall the exact form of the Kerr–NUT–(anti-)de Sitter metric tensor and we briefly remind the problems those spacetimes have.
In the following sections an invariant, coordinate independent approach will be introduced.

The Kerr–NUT–(anti-)de Sitter spacetime is defined by the following metric tensor \cite{griffiths_podolsky_2009}
\begin{equation}
\begin{split}
\label{eq:MetricKNdS}
    ds^2=-\frac{\mathcal{Q}}{\Sigma}(dt-A d\phi)^2  +\frac{\Sigma}{\mathcal{Q}}dr^2 
+\frac{\Sigma}{P}d\theta^2+\frac{P}{\Sigma}\sin^2\theta(adt-\rho d\phi)^2,
\end{split}
\end{equation}
where
\begin{equation}
    \label{eq:functions-metric}
    \begin{split}
        \Sigma&=r^2+(l+a\cos\theta)^2,\\
        A&=a\sin^2\theta+4l\sin^2\tfrac{1}{2}\theta,\\
        \rho&=r^2+(l+a)^2=\Sigma+aA,\\
        \mathcal{Q}&=(a^2-l^2)-2mr+r^2-\Lambda\big((a^2-l^2)l^2+(\tfrac{1}{3}a^2+2l^2)r^2+\tfrac{1}{3}r^4\big),\\
        P&=1+\frac{4}{3}\Lambda al  \cos\theta+\frac{\Lambda}{3}a^2\cos^2\theta.
    \end{split}
\end{equation}
It satisfies the vacuum Einstein equations with a cosmological constant
\begin{equation}
    G_{\mu\nu}+\Lambda g_{\mu\nu}=0.
\end{equation}
This solution is commonly interpreted as a rotating black hole on the (anti-)de Sitter background, additionally equipped with a NUT parameter $l$. 
The NUT parameter can be seen as a gravitational analogue of the magnetic monopole, introducing a topological defect. 
In fact, metrics with $l\neq0$ are known to posses a string-like conical singularity.
Indeed, if a surface $t,r=$ const is assumed to be diffeomorphic to sphere and the coordinates $\theta$ and $\phi$ are assumed to be the spherical coordinates, then a non-continuity is caused by the non-vanishing of the $1$-form $Ad\phi$ at the pole $\theta=\pi$.
Obviously this is a coordinate dependent statement, and a transformation 
\begin{equation}
    \label{eq:transformation_change_of_singular_axis}
    t'=t-4l\phi
\end{equation}
transforms $A$ into $A'=a\sin^2\theta-4l\cos^2\tfrac{1}{2}\theta$ that vanishes at $\theta=\pi$. However, $A'$ does not vanish at $\theta=0$. 
Another drawback of the coordinate $t'$ is a discontinuity on the would be circles parametrised by $\phi$.
These observations lead to a possibility of gluing halves of the two non-singular coordinate patches, and making $t$ a cyclic variable parametrising circles.
In consequence, the spacetime topology takes the form (\ref{eq:bundle}). 
In the case $a=0$ that topological step is sufficient, and it removes all the singularities from the axis $\theta = 0,\pi$.
In the $l, a\not=0$ case however, for generic values of the parameters $(m,a,l,\Lambda)$ a non-removable non-differentiability persists at least either at the $\theta=0$ or at the $\theta=\pi$ half-axis.
It is an irremovable obstacle, except for a special, three dimensional subclass of the Kerr-NUT-de Sitter spacetimes we derive below in Section \ref{sec:non-singular-KNDS}. 

Another seemingly singular element of the metric tensor (\ref{eq:MetricKNdS}) is every zero of the function $\mathcal{Q}$.
But those become regular null surfaces after suitable coordinate change, and give rise to familiar Killing horizons, which we use intensively in this paper.
In general $\mathcal{Q}$ has up to four roots, each of them defines an inner black hole horizon, an outer black hole horizon or a cosmological horizon.
Finally, the presence of the NUT parameter $l$ in the formula for the function $P$ allows that function to change a sign unless $l$ is suitably bounded.
The change of the sign would cause the change of the signature of the metric tensor (\ref{eq:MetricKNdS}), hence it could not be compensated by any transformation of coordinates.
In particular, if we are interested in the spacetimes of the signature $(-+++)$, $P\ge 0$ everywhere is the necessary condition. Thus we restrict the parameter space $(\Lambda,a,l)$ accordingly, to the following subset
\begin{equation}
\label{eq:P-ineq}
       P>0 \iff \Bigg(\bigg( 2\bigg|\frac{l}{a}\bigg|\leq 1 \land \Lambda < \frac{3}{4l^2}\bigg) \lor  \bigg( 2\frac{l}{a}\geq 1 \land \Lambda < \frac{-3}{a^2-4al}\bigg) \lor  \bigg(-2\frac{l}{a}\geq 1 \land \Lambda < \frac{-3}{a^2+4al}\bigg)\Bigg)
\end{equation}

\section{Singular and non-singular horizons in Kerr–NUT–(anti-)de Sitter}
\label{sec:horizonKNUT}
\subsection{The horizons}\label{subsec:horizon}
The metric (\ref{eq:MetricKNdS}) is obviously not well defined at the surfaces of $r=r_0$, where $r_0$ is a root of $\mathcal{Q}$. 
To remediate this we can introduce an advanced null coordinate $v$ and a new angular coordinate $\tilde{\phi}$ such that associated coframe is
\begin{equation}
\begin{split}
\label{eq:coframe-horizon-non-singular}
    dv=dt+\frac{\rho}{\mathcal{Q}}dr,\\
    d\Tilde{\phi}=d\phi +\frac{a}{\mathcal{Q}}dr.
    \end{split}
\end{equation}
Then the metric tensor (\ref{eq:MetricKNdS}) takes the following form
\begin{equation}
\label{eq:metric-non-singular}
    ds^2=-\frac{\mathcal{Q}}{\Sigma}(dv-A d\Tilde{\phi})^2+2 dr (dv-A d\Tilde{\phi})+\frac{\Sigma}{P}d\theta^2+\frac{P}{\Sigma}\sin^2\theta(a dv-\rho d \Tilde{\phi})^2.
\end{equation}
Now the above metric is non-singular when $\mathcal{Q}(r)=0$ and the surfaces $r=r_0$ are null hypersurfaces of induced degenerate geometry 
\begin{equation}
\label{horizon}
 {}^{(2)}q_H  = \frac{\Sigma_0}{P}d\theta^2+\frac{P}{\Sigma_0}\sin^2\theta(a dv-\rho_0 d \Tilde{\phi})^2,
\end{equation}
where 
\begin{equation}
\Sigma_0(\theta):=\Sigma(r_0,\theta), \qquad \rho_0 := \rho(r_0). 
\end{equation} 

Their geometric importance becomes clear if we consider the algebra of Killing vector fields of the metric (\ref{eq:metric-non-singular}), spanned by commuting vector fields $\partial_v$ and $\partial_{\phi}$. 
All of them are tangent to the $r=r_0$ surface. 
Moreover, the vector field 
\begin{equation}
    \xi=\partial_v+\Omega_H \partial_{\tilde{\phi}},
\end{equation}
where 
\begin{equation}
    \Omega_H:=\frac{a}{\rho_0}=\frac{a}{r^2_0+(a+l)^2},
\end{equation}
becomes null thereon,
\begin{equation}
\label{eq:killing-norm}
    g(\xi,\xi)=-\frac{\mathcal{Q}}{\Sigma}(1-A \Omega_H)^2+\frac{P}{\Sigma}\sin^2\theta a^2(1-\frac{\rho}{\rho_0})^2 \bigg|_{r=r_0}=0.
\end{equation}
Thus we conclude that for every root $r_0$ of the function $\mathcal{Q}$, the surface $r=r_0$ is a Killing horizon associated with the corresponding Killing vector $\xi$. 

\subsection{The space of the null generators and its geometry}
In this subsection we investigate the geometry of the null generators of the horizon. This a geometrically defined structure hence its properties lead us to conclusions on the metric tensor (\ref{eq:MetricKNdS}) independent of the choice of the coordinates $(t,r,\theta,\phi)$. 

First, we introduce two coordinates 
$(x^2,x^3)$ on the horizon, constant along the null generators,
\begin{equation*}
\xi(x^i)=0, \ \ \ \ \ \ i=2,3.
\end{equation*}
They can also serve as coordinates on the moduli space.
On the other hand, completed by the function $\tau$ such that $\xi(\tau)\neq0$, they will set a new coordinate system on the horizon. Subsequently we define
 \begin{equation}
 \label{eq:killing-coords-horizon}
\begin{bmatrix}
\tau\\
x^2\\
x^3
\end{bmatrix}
:=
\begin{bmatrix}
v\\
\theta\\
-\Omega_Hv+ \tilde{\phi}
\end{bmatrix}.
\end{equation}
Then the corresponding tangent frame is
\begin{equation}
\label{eq:frame-killing-coords-horizon}
\begin{bmatrix}
\partial_\tau\\
\partial_2\\
\partial_3
\end{bmatrix}
=
\begin{bmatrix}
\xi\\
\partial_\theta\\
 \partial_{\tilde{\phi}}
\end{bmatrix}.
\end{equation}
In the new coordinates the (degenerate) horizon metric tensor (\ref{horizon}) becomes

\begin{equation}
\label{eq:metric-horizon}
        {}^{(2)}q_H=\frac{\Sigma_0}{P}(dx^2)^2+ \frac{P}{\Sigma_0}\sin^2(x^2) \rho_0^2 (dx^3)^2.
\end{equation}
Without prejudging the spherical nature of the variables $(x^2,x^3)$ we can say, that as long as
\begin{equation*}
x^2\not=0,\pi,
\end{equation*}
 ${}^{(2)}q_H$ is an analytic metric tensor of Riemannian signature.
 On the other hand, the equalities 
\begin{equation}
\label{eq:q33}
q_{33}(0,x^3) = 0 = q_{33}(\pi,x^3)
\end{equation} 
tell us that the zero sets of $x^2$ and $x^2-\pi$ are two single points.
We call them poles. 
The 
\begin{equation*}
    x^3=\rm const
\end{equation*}
curves emanate from the pole $x^2=0$ to reach the pole $x^2=\pi$, and are geodesic. 

That means that the surfaces of constant $x^2$ are circles of equal geodesic distance centred around the poles.
Hence they are closed loops, parametrised by $x^3$ taking values in a fixed interval, say
\begin{equation*}
\label{eq:interval}
x^3 \in [0, 2\pi c), \ \ \ \ \ \ c=\rm const. 
\end{equation*}
The interval has to be independent of $x^2$, because otherwise there would be points not connectable with one of the poles with a geodesic curve. 

Away from the poles, the tensor ${}^{(2)}q_H$ is analytic and defines a Riemannian metric tensor.
If the space and the metric is extendable to the poles in a differentiable way, then the circumference $L(x^2)$ of a circle $x^2=\text{const}$ and the length $R_0(x^2)$/$R_\pi(x^2)$ of a geodesic segment from the pole $x^2=0$ or $x^2=\pi$ to  the chosen value of $x^2$ have to satisfy the following boundary conditions:
\begin{equation}
\label{theconstraint}
\lim_{x^2\rightarrow 0}\frac{L(x^2)}{R_0(x^2)} = 2\pi = \lim_{x^2\rightarrow \pi}\frac{L(x^2)}{R_\pi(x^2)} .
\end{equation}

\subsection{The necessary singularity removability condition}
Now, we calculate the limits (\ref{theconstraint}) and derive the singularity removability condition. 
The circumference is defined as
\begin{equation}
    L(x^2)=\int_0^{2\pi c}\sqrt{q_{33}} dx^3.
\end{equation} 
The radius is calculated differently in the neighbourhood of the different poles: \\ 
for $0<x^2\leq \pi/2$ we use
\begin{equation}
    R_0(x^2)=\int_0^{x^2}\sqrt{q_{22}} dx^2
\end{equation}
and for $\pi/2 < x^2 < \pi$ we use
\begin{equation}
    R_\pi(x^2)=\int^\pi_{x^2}\sqrt{q_{22}} dx^2.
\end{equation}
Then the limit of the ratio is
\begin{equation}
\begin{split}
    \lim_{x^2 \to 0/\pi} \frac{L(x^2)}{R_{0/\pi}( x^2)}&=\lim_{ x^2 \to 0/\pi} 2\pi c {\Sigma_0^{-1/2 }P^{1/2}} \bigg(\partial_{ x^2}\Big(\rho_0\Sigma_0^{-1/2}P^{1/2}\Big)\sin x^2 + \Big(\rho_0\Sigma_0^{-1/2}P^{1/2}\Big)\cos x^2\bigg)=\\
    &=2\pi c  \frac{\rho_0 P}{\Sigma_0}\bigg|_{ x^2=0/\pi}=
\begin{dcases} 
      2\pi c P(0) & \text{for }  x^2=0, \\
      2\pi c \frac{r_0^2+(l+a)^2}{r_0^2+(l-a)^2} P(\pi) & \text{for }  x^2=\pi.
   \end{dcases}
  \end{split}
\end{equation}

We can see that $c$ can be set to a value reproducing $2\pi$ at the both poles if and only if the following condition is satisfied
\begin{equation}
\label{eq:condition-for-diff-horizon}
    P(0)=\frac{r_0^2+(l+a)^2}{r_0^2+(l-a)^2} P(\pi).
\end{equation}
If that is the case, then
\begin{equation}
\label{cfixed}
  c=  \frac{1}{P(0)}.
\end{equation}

\subsection{Kerr–NUT–anti-de Sitter spacetimes admitting a non-singular horizon}
The equality (\ref{eq:condition-for-diff-horizon}) is a condition on the parameters $(r_0,a,l,\Lambda)$.
The solutions to this conditions come in three (not exclusive) classes:
\begin{itemize}
    \item $l=0$, $a,\Lambda \in \mathbb{R}$
    \item $a=0$, $l,\Lambda \in \mathbb{R}$
    \item $\Lambda a l \not= 0$. The value of $\Lambda$ is determined by the condition (\ref{eq:condition-for-diff-horizon}) as a function of $(r_0,a,l)$ and evaluates to
    \begin{equation} 
\label{eq:constraint-Lambda}
  \Lambda =  \frac{3}{a^2+2l^2+2r_0^2}.
  \end{equation}
  From now on we focus on the solutions characterised by (\ref{eq:constraint-Lambda}). 
\end{itemize}
Note that the $\Lambda$ is necessarily positive excluding Kerr–NUT–anti-de Sitter spacetimes. Given $(a,l)$ and a root $r_0$ of the function $\mathcal{Q}$, the mass parameter $m$ is determined by the identity following from the vanishing of the function $\mathcal{Q}$
 \begin{equation}
\label{eq:mass-constraint}
\begin{split}
m =& \frac{1}{2r_0}\Big( ( a^2-l^2)+r_0^2-\Lambda\big((a^2-l^2)l^2+(\tfrac{1}{3}a^2+2l^2)r_0^2+\tfrac{1}{3}r_0^4\big)\Big)=\\
=&\frac{a^4 - 2 a^2 l^2 + l^4 + 2 a^2 r_0^2 - 6 l^2 r_0^2 + r_0^4}{2 a^2 r_0 + 4 l^2 r_0 + 4 r_0^3}.
\end{split}
\end{equation}
It is immediately obvious from the last equality of the above formula that the sign of $m$ can be appropriately chosen by changing the sign of $r_0$ (and only $r_0$, the sign of $m$ is insensitive to $a$ and $l$).

\textcolor{black}{Another immediate consequence of the (\ref{eq:constraint-Lambda}) is that for a given choice of the parameters defining the spacetime, at most one horizon (precisely the one the radius $r_0$ satisfying (\ref{eq:constraint-Lambda})) can be made non-singular.}
\subsection{The resulting horizon geometry}
We assume through out this paper, that the function $P$ nowhere vanishes on the space of the null generators of the horizon.
A good news is that actually every sequence of the parameters $(m,a,l,\Lambda)$ that satisfy the singularity removability conditions (\ref{eq:constraint-Lambda}) automatically falls in one of the three cases on the RHS of (\ref{eq:P-ineq}). 

By assuming the necessary singularity removability conditions, we determined a 3-parameter family of geometries ${}^{(2)}q_H$ defined on a topological 2-sphere.
The sphere is parametrised by the variables 
\begin{equation*}
   (x^2,x^3)\in [0,\pi]\times\left[0, \frac{2\pi}{P(0)}\right). 
\end{equation*}
Knowing the exact domain of the variables we can calculate the $2$-area of the moduli space of the null generators (see Sec. \ref{sec:comparison}).
The result is very simple, namely from  (\ref{eq:area-2-form}) we have
\begin{equation}
   \text{Area}= 4\pi \left(\frac{3}{2\Lambda} \right) .
\end{equation}
If we assume that the variables $(x^2, P(0) x^3)$ set a spherical coordinate system on $S^2$ customarily denoted by $(\theta,\varphi)$, then we can check by inspection (see Appendix \ref{sec:appendix}), that the metric tensor ${}^{(2)}q_H$ is smooth at the poles. 

Our result has a tricky consequence.
Suppose that a Kerr-NUT-de Sitter spacetime (\ref{eq:MetricKNdS}) is defined by parameters $(m,a,l,\Lambda)$ such that for one of the horizons of a radius $r_0$ the singularity removability condition (\ref{eq:constraint-Lambda}) for $\Lambda\not=0$ is satisfied.
Then, any other Killing horizon contained in that spacetime has a radius $r_0' \neq r_0$, and the singularity removability condition (\ref{eq:constraint-Lambda}) can not be satisfied (with $r'_0$ substituted for $r_0$) any more. 

While the space of the null generators is $S^2$ endowed with the metric tensor ${}^{(2)}q_H$, the horizon itself is not just $S^2\times \mathbb{R}$.
A geometric approach to that problem is to consider the rotation-connection $1$-form $C$ (a name made up for this paper).
Anticipating the results from Sec. \ref{sec:hopf} we define $C$ to be
\begin{equation}
    C=\frac{\xi_\mu}{g(\xi,\xi)}  dx^\mu.
\end{equation}
The pull back to the horizon $r=r_0$ (we are assuming $r_0$ is a single root of $\mathcal{Q}$) takes the following form
\begin{equation}
C\Big|_{r=r_0} = d\tau - \frac{A \rho_0}{\Sigma_0}dx^3.
\end{equation}
It is not well defined on the entire $S^2$ parametized by $(x^2,x^3)$, as the part $-\frac{A \rho_0}{\Sigma_0}dx^3$ is not well defined at $ x^2=\pi$.
We go back to this issue in Sec. \ref{sec:hopf}

\section{The Killing orbit space in neighbourhood of non-singular horizon}

In this section we investigate a future/past neighbourhood of a non-singular horizon in Kerr-NUT-de Sitter spacetime. 
To that end we consider the Killing vector field $\xi$ that develops the horizon and derive the geometry induced on the 3-dimensional space of orbits of $\xi$ near the horizon. 
General considerations concerning this approach can be found here \cite{chrusciel}.
We discover that, somewhat magically, the horizon singularity removability condition (\ref{eq:condition-for-diff-horizon}) implies the singularity removability in the neighbourhood.
We need to make a new assumption, though.
Thus far, in the horizon geometry case, the multiplicity of a root $r_0$ of the function $\mathcal{Q}$ was not relevant.
It turns out, however, that now the multiplicity of the root plays a role.
For simplicity, we will focus on the non-degenerate (non-extremal) case, that is assume that the root $r_0$ is single.
We will aware the reader when this assumption is used. 

Clearly, the $3$-dimensional orbit space geometry is degenerate wherever 
\begin{equation*}
\xi^\mu\xi_\mu=0.    
\end{equation*}
Hence, we consider a region in spacetime such that 
\begin{equation}
\label{eq:eps}
r_0<r<r_0 + \epsilon, \ \ \ \ \ \ \ {\rm or} \ \ \ \ \ \ \ r_0 - \epsilon<r< r_0, 
\end{equation}
where $\epsilon>0$ is sufficiently small to ensure \begin{equation*}
  \xi^\mu\xi_\mu\neq0.  
\end{equation*}
As can be seen from (\ref{eq:killing-norm}), close to the horizon the sign of the $\xi^\mu\xi_\mu$ is governed by the sign of $\mathcal{Q}$ such an $\epsilon$ always exists.
Similarly, if the horizon is non-extremal and $r_0$ is a single root of the polynomial $\mathcal{Q}$ then $\xi$ is strictly timelike/spacelike on one side of the horizon, and strictly spacelike/timelike on the other side.

\subsection{Derivation of the induced geometry}
Similarly to the previous section, in order to introduce coordinates on the space of the orbits of $\xi$, we need three independent functions $(x^1,x^2,x^3)$ in spacetime, such that 
\begin{equation}
\xi(x^i)=0, \ \ \ \ \ \ i=1,2,3.
\end{equation}
Since we are away from the zeros of $\mathcal{Q}$, as a starting point we may use the spacetime coordinates $(t,r,\theta,\phi)$ and the spacetime metric (\ref{eq:MetricKNdS}).
We introduce a new system of spacetime coordinates
 \begin{equation}
 \label{eq:killing-coords}
\begin{bmatrix}
\tau\\
x^1\\
x^2\\
x^3
\end{bmatrix}
:=
\begin{bmatrix}
t\\
r\\
\theta\\
-\Omega_H t+ \phi
\end{bmatrix},
\end{equation}
with the associated tangent frame
\begin{equation}
\label{eq:frame-killing-coords}
\begin{bmatrix}
\partial_\tau\\
\partial_1\\
\partial_2\\
\partial_3
\end{bmatrix}
=
\begin{bmatrix}
\xi\\
\partial_r\\
\partial_\theta\\
 \partial_{\phi}
\end{bmatrix}.
\end{equation}
In the new coordinates the metric (\ref{eq:MetricKNdS}) takes the following form 
\begin{equation}
\label{eq:metric-knds-killing-variables}
    ds^2=-\frac{\mathcal{Q}}{\Sigma}\big(\frac{\Sigma_0}{\rho_0}d\tau-A dx^3\big)^2  +\frac{\Sigma}{\mathcal{Q}}(dx^1)^2 +\frac{\Sigma}{P}(dx^2)^2+\frac{P}{\Sigma}\sin^2 (x^2)\Big(\frac{a}{\rho_0}\big((x^1)^2-r_0^2) d\tau-\rho dx^3\Big)^2,
\end{equation}
 where we used a fact that $1-A\Omega_H=\Sigma_0/\rho_0$. The coefficients of the metric that require a short calculation are:
\begin{equation}
\label{eq:g_coef}
\begin{split}
    &g_{\tau\tau}=-\frac{\mathcal{Q}}{\Sigma}\frac{\Sigma_0^2}{\rho_0^2}+\frac{P}{\Sigma}\sin^2 (x^2) \frac{a^2}{\rho_0^2}((x^1)^2-r_0^2)^2,\\
    &g_{33}=-\frac{\mathcal{Q}A^2}{\Sigma}+\frac{P}{\Sigma}\sin^2 (x^2) \rho^2,\\
    &g_{\tau 3}=\frac{\mathcal{Q}}{\Sigma}\frac{\Sigma_0}{\rho_0}A+ \frac{P}{\Sigma}\sin^2 (x^2) \frac{a}{\rho_0} \rho ((x^1)^2-r_0^2).
\end{split}
\end{equation}
With some harmless abuse of notation the functions $x^i$ (\ref{eq:killing-coords}) constitute a coordinate system on the orbit space of the Killing vector field $\xi$.
The metric $q$ induced on the space of orbits measures the spacetime distance between orbits in a frame in rest with respect to an observer following a given orbit of $\xi$ (that is if $\xi$ is timelike; however the formula below makes also perfect geometric sense for a spacelike $\xi$), then
\begin{equation}
    q_{ij}:=g(\hat{\partial_{i}},\hat{\partial_{j}}),
\end{equation}
where the vector fields $\hat{\partial}_i$ are defined to satisfy the following conditions:
\begin{equation}
\label{eq:def-hat-frame}
\hat{\partial_{i}}=\partial_i-A_i \xi, \ \ \ \ {\rm and} \ \ \ \  g(\xi,\hat{\partial}_i)=0. 
\end{equation}
It is easy to see that 
\begin{equation}
    A_i=\frac {g_{i\tau}}{g_{\tau\tau}}
\end{equation}
and hence 
\begin{equation}
    q_{ij}=g_{ij}-\frac {g_{i\tau}g_{j\tau}}{g_{\tau\tau}}.
\end{equation}

Finally, the metric $q$ of the moduli space is
\begin{equation}
\label{eq:metric-orbits-KNDS}
    q=\frac{\Sigma}{\mathcal{Q}}(dx^1)^2+\frac{\Sigma}{P}(dx^2)^2+\frac{P \mathcal{Q}\sin^2(x^2) \rho_0^2\Sigma}{\mathcal{Q}\Sigma_0^2- P \sin^2(x^2) a^2((x^1)^2-r_0^2)^2} (dx^3)^2.
\end{equation}

\subsection{The signature and non-differentiable points}
For the values of $x^1$, simillarly to (\ref{eq:eps}) we consider either
\begin{equation}
r_0<x^1<r_0 + \epsilon, \ \ \ \ \ \ \ {\rm or} \ \ \ \ \ \ \ r_0 - \epsilon<x^1< r_0, 
\end{equation}
and then for 
$$x^2\not=0,\pi$$ 
every component $q_{ij}$ is differentiable, even smooth. 
Indeed, $\mathcal{Q}(x^1)\not=0$ makes $q_{11}$ a non-vanishing and smooth function.
The component $q_{22}$ is smooth and differentiable everywhere by the assumption $P>0$.
The denominator of $q_{33}$ does not vanish as long as $g_{\tau\tau}$ does not, that is as long as $\xi$ is not null, hence $q_{33}$ is smooth too. 

The signature of the metric tensor $q$ is $(+++)$ if $\xi$ is timelike, and $(-++)$ if $\xi$ is spacelike.
To see that, we study the signs of the functions $q_{11}, \,q_{22}$ and $q_{33}$.
Clearly, $q_{22}>0$ everywhere.
The component $q_{11}$ can be written as
\begin{equation}
    q_{11}=(-\frac{\Sigma^2_0}{\rho_0^2}-P\sin^2(x^2)\frac{a^2}{\rho^2_0}\frac{((x^1)^2-r_0^2)^2}{\mathcal{Q}})g(\xi,\xi)^{-1}.
\end{equation}
For an $\epsilon>0$ sufficiently close to the non-extremal horizon the sign of $q_{11}$ is determined by the sign of $g(\xi,\xi)$ and $q_{11}$ is positive for timelike $\xi$ and negative for spacelike $\xi$.
The component $q_{33}$ looks more complicated, however again for a sufficiently small $\epsilon$ in (\ref{eq:eps}), we can estimate it by its limit at $r=r_0$, which is
\begin{equation}
\label{limit}
\lim_{r\rightarrow r_0} q_{33}\ =\    \frac{P \sin^2(x^2) \rho_0^2}{\Sigma} >0 .  \end{equation}   
Here the assumption that $r_0$ is a single root is used.
Indeed, 
\begin{equation}
    q_{33}= \frac{P \sin^2 (x^2) \rho_0^2\Sigma}{\Sigma_0^2- P \sin^2 (x^2) a^2(x^1+r_0)^2\frac{(x^1-r_0)^2}{\mathcal{Q}}}, 
\end{equation}
and 
\begin{equation}\label{limitb}
\lim_{x^1\rightarrow r_0} \frac{(x^1-r_0)^2}{\mathcal{Q}} = 0,
\end{equation}
provided $r_0$ is a single root of $\mathcal{Q}$.
Otherwise, the limit is finite but not zero if the root is double, and even infinite if the root is triple. 

Notice, that in the single root case,
\begin{equation}
\label{limita}
\lim_{x^1\rightarrow r_0} \left(q_{22}(dx^2)^2+q_{33}(dx^3)^2\right)\ =\  {}^{(2)}q_H,  
\end{equation}  
where the right hand side is the horizon null generators geometry of the previous section.
In a higher multiplicity case, the left hand side of (\ref{limitb}) changes the limit.

The only possible non-differentiability of the metric $q$ (\ref{eq:metric-orbits-KNDS}) is that of the metric tensor 
\begin{equation}
{}^{(2)}q:=\frac{\Sigma}{P}(dx^2)^2+\frac{P \mathcal{Q}\sin^2(x^2) \rho_0^2\Sigma}{\mathcal{Q}\Sigma_0^2- P \sin^2(x^2) a^2((x^1)^2-r_0^2)^2} (dx^3)^2
\end{equation}
defined for every value of 
$$x^1={\rm const}$$
on the manifold diffeomorphic to $S^2$, and the only potentially non-differentiable points are 
\begin{equation*}
  x^2=0,\pi.  
\end{equation*}
We will study them in the next subsection by applying the same method as in the previous section.

\subsection{The differentiability condition}
\label{sec:non-singular-KNDS}
 Now, we turn to the analysis of the properties of the metric tensor ${}^{(2)}q$ induced by (\ref{eq:metric-orbits-KNDS}) on a surface $x^1=r$.
 We proceed with studying the differentiability condition in an analogous manner as in the horizon case, in particular we apply the same notation for the circumference $L( x^2)$ and radius $R_0( x^2)$/$R_\pi( x^2)$ of circles around the poles.
 The conditions are again that the ratio of the circumference to the radius of the circles of $x^2=\theta$ shrunk to a pole approaches $2\pi$.

\begin{equation}
\begin{split}
    &\lim_{ x^2 \to 0/\pi} \frac{L( x^2)}{R_{0/\pi}( x^2)}=\lim_{ x^2 \to 0/\pi} 2\pi c {\Sigma^{-1/2 }P^{1/2}} \Bigg(\partial_{ x^2}\bigg(\frac{P^{1/2} \mathcal{Q}^{1/2}\rho_0\Sigma^{1/2}}{\big(\mathcal{Q}\Sigma_0^2- P \sin^2( x^2) a^2((x^1)^2-r_0^2)^2\big)^{1/2}}\bigg)\sin( x^2) + \\
    &+\bigg(\frac{P^{1/2} \mathcal{Q}^{1/2}\rho_0\Sigma^{1/2}}{\big(\mathcal{Q}\Sigma_0^2- P \sin^2( x^2) a^2((x^1)^2-r_0^2)^2\big)^{1/2}}\bigg)\cos( x^2)\Bigg)=2\pi c  \frac{\rho_0 P}{\Sigma_0}\bigg|_{x^2=0/\pi}=
\begin{dcases} 
      2\pi c P(0) & \text{for } x^2=0, \\
      2\pi c \frac{r_0^2+(l+a)^2}{r_0^2+(l-a)^2} P(\pi) & \text{for } x^2=\pi.
   \end{dcases}
  \end{split}
\end{equation}
Somewhat unexpectedly, we reach the same condition for removing the singularities as for the horizon (\ref{eq:condition-for-diff-horizon}):
\begin{equation}
\label{eq:condition-for-diff}
    P(0)=\frac{r_0^2+(l+a)^2}{r_0^2+(l-a)^2} P(\pi).
\end{equation}
Notice, that even though the calculation does depend on the fixed value $r_0$ of the coordinate $x^1$, the result is $x^1(=r)$ independent.
Hence, it is satisfied simultaneously for all the values of $r$, even without restriction by (\ref{eq:eps}).
Also, the value of the rescaling parameter $c$ necessary to remove the singularity, namely
\begin{equation}
\label{eq:scaling-const}
 c=\frac{1}{P(0)}  = \frac{1}{1+\frac{4}{3}\Lambda al  +\frac{\Lambda}{3}a^2}=\frac{3}{2\Lambda \rho_0}
\end{equation}
is valid for every value of $r$.

\textcolor{black}{Here, as in the horizon case, the neighbourhood of at most one of the horizons can be made non-singular.
This is a worrisome feature, yet there are still open questions offering a possible way out.
If we can control which horizon is made non-singular and extend the non-singular neighbourhood through the (physically uninteresting) surfaces of vanishing norm of the Killing vector, which are not horizons, our solutions would have two immediate, possible applications. 
In the first scenario the most exterior horizon is non-singular and we can get some insight about the asymptotic behaviour of the non-singular Kerr-NUT-de Sitter spacetimes.
In the second scenario, the second most exterior horizon is non-singular and the spacetime could be used to model the exterior of some exotic black hole.}

Thus far we have examined the necessary conditions for the removability of the conical singularity in the metric tensor (\ref{eq:metric-orbits-KNDS}) induced on the space of orbits of the Killing vector $\xi$ developing a non-extremal Killing horizon, and we found they are satisfied if and only the horizon itself satisfies the removability condition.
On the other hand, if we assume that the coordinate system $(x^1,x^2,{x^3}{P(0)})$ is a spherical coordinate system in $\mathbb{R}^3$ usually denoted by the $(r,\theta,\varphi)$, then we can check by inspection (Appendix \ref{sec:appendix}) that upon the condition (\ref{eq:constraint-Lambda}) the metric tensor $q$ actually is smooth at the poles.

\section{The Hopf bundle extension} 
\label{sec:hopf}
The goal of this section is to reconstruct a conical singularity free spacetime geometry 
of the singularity free orbit space geometry $q$ (\ref{eq:metric-orbits-KNDS}) derived above. 

Given a metric tensor $g$ and a Killing vector $\xi$ such that 
\begin{equation*}
  \xi^\mu\xi_\mu\not=0,  
\end{equation*}
the fibration structure defined by $\xi$ is endowed with the following geometric structures: geometry of the space of the orbits, the rotation-connection $1$-form 
$\frac{1}{g(\xi,\xi)}\xi_\mu dx^\mu$ and the laps function $\xi^\mu\xi_\mu$. To characterise them we choose coordinates $(\tau,x^i)$ such that
\begin{equation*}
    \xi=\partial_\tau.
\end{equation*}
Then we write \cite{chrusciel}
\begin{equation}
\begin{split}
g = &g_{\tau\tau}d\tau^2 + 2 g_{\tau i}d\tau dx^i + g_{ij}dx^idx^j= g_{\tau\tau}\left(d\tau + \frac{g_{\tau i}}{g_{\tau \tau}}dx^i \right)^2 - \frac{g_{\tau i}g_{\tau j}}{g_{\tau\tau}}dx^idx^j + g_{ij}dx^idx^j=\\
=&g_{\tau\tau}(d\tau + \omega_i dx^i)^2 + q_{ij}dx^idx^j . 
\end{split}
\end{equation} 
Now, $q_{ij}dx^idx^j$ is the orbit space metric tensor, $dt +\omega_idx^i$ is the rotation-connection 1-form, and the lapse function is $\xi^\mu\xi_\mu=g_{\tau\tau}.$ 
The advantage of that approach is that the geometric structures have a coordinate independent meaning and should be well defined. 
We apply it to the metric tensor (\ref{eq:metric-knds-killing-variables}) and the Killing vector field $\xi$. 
The orbit space metric tensor is $q$ (\ref{eq:metric-orbits-KNDS}).
We write it below in terms of the partially rescaled angular coordinates
\begin{equation}
(r,\theta,\varphi):=(x^1,x^2,{P(0)}{x^3}).
\end{equation}
They are now legitimate spherical coordinates in $\mathbb{R}^3$ and the $3$-metric tensor is 
\begin{equation}
\label{eq:metric-neighbourhood-nonsingular}
    q=\frac{\Sigma}{\mathcal{Q}}dr^2+\frac{\Sigma}{P}d\theta^2+\frac{P \mathcal{Q}\sin^2\theta \rho_0^2\Sigma}{\mathcal{Q}\Sigma_0^2- P \sin^2\theta a^2(r^2-r_0^2)^2} \frac{1}{P^2(0)}d\varphi^2,
\end{equation}
It is defined on
\begin{equation}\label{RS2}
(r_0-\epsilon,r_0)\times S^2\ \cup\ (r_0,r_0+\epsilon,)\times S^2,
\end{equation}
and is everywhere differentiable provided the singularity removability condition (\ref{eq:condition-for-diff-horizon}) is satisfied.
The spacetime metric tensor (\ref{eq:metric-knds-killing-variables}) adapted to the orbit fibration structure of the spacetime induced by the Killing vector $\xi=\partial_\tau$ takes the following form
\begin{equation}
\label{eq:ds-metric}
ds^2 = g_{\tau\tau}\left(d\tau + \frac{g_{\tau 3}}{g_{\tau\tau}} dx^3\right)^2 + q = g_{\tau\tau}\left(d\tau + \frac{g_{\tau 3}}{g_{\tau\tau}} \frac{d\varphi}{P\left(0\right)}\right)^2 + q
\end{equation}
However, the part of the rotation $1$-form 
\begin{equation} 
\label{eq:rot-1-form}
\frac{g_{\tau 3}}{g_{\tau\tau}{P\left(0\right)}} {d\varphi}=:\omega d\varphi
\end{equation}
is not well defined at the pole $p_\pi\in S^2$ corresponding to
\begin{equation}
\label{eq:pi}
\theta=\pi.
\end{equation}
Indeed, the obstacle, for every value of $r$, is non-vanishing of the coefficient $\omega(r,\pi)$.
Notice however, that the value is $r$ independent, indeed
\begin{equation}
\omega(\pi) := \omega(r,\pi)= -\frac{4l}{P(0)}\frac{r_0^2 + (a+l)^2}{r_0^2 + (a-l)^2}= -\frac{4l}{P(\pi)} = -\frac{4l}{1-\frac{4}{3}\Lambda al +\frac{\Lambda}{3}a^2} \not=0 , 
\end{equation}
where we used the singularity removability condition (\ref{eq:condition-for-diff-horizon}) to replace $P(0)$ by $P(\pi)$. 
To cover the points (\ref{eq:pi}), we define another chart $(\tau',r',\theta',\varphi')$ related to $(\tau,r,\theta,\varphi)$ by the following coordinate transformation, 
\begin{equation}
\label{eq:trans}
\tau = \tau'-\omega(\pi)\phi',\ \ \ r=r', \ \ \ \theta=\theta',\ \ \ \varphi = \varphi' . 
\end{equation} 
The resulting metric reads
\begin{equation}
\label{eq:ds'-metric}
ds'^2 = g_{\tau\tau}(r',\theta')\left(d\tau' + \left(\omega\left(r',\theta'\right)-\omega\left(r,\pi\right)\right) d\varphi'\right)^2 + q_{rr}(r',\theta')(dr')^2 + q_{\theta\theta}(r',\theta')(d\theta')^2+ q_{\varphi\varphi}(r',\theta')(d\varphi')^2.
\end{equation}
It is well defined except for the pole $p_0\in S^2$, where
\begin{equation*}
\theta=0 .
\end{equation*}
To make the transformation (\ref{eq:trans}) well defined for every value of $\tau$, we have to assume that $\tau$ is a cyclic variable that parametrises a circle $S^1$ and ranges the interval
\begin{equation*}
\tau \in \left[0, 2\pi\omega(\pi)\right[.
\end{equation*}
The global structure of the resulting spacetime is as follows:
The metric tensor $ds^2$ (\ref{eq:ds-metric}) is defined on
\begin{equation}
S^1\times \left((r_0-\epsilon,r_0) \cup\ (r_0,r_0+\epsilon)\right)\times \left(S^2\setminus \{p_\pi\}\right),
\end{equation}
covered by the coordinate system $(\tau,r,\theta,\phi)$, while the metric tensor $ds'^2$ (\ref{eq:ds'-metric}) is defined on
\begin{equation}
S^1\times \left((r_0-\epsilon,r_0) \cup\ (r_0,r_0+\epsilon)\right)\times \left(S^2\setminus \{p_0\}\right),
\end{equation}
covered by the coordinate system $(\tau',r',\theta',\phi')$ and the transformation (\ref{eq:trans}) between the coordinates is defined on 
\begin{equation}
S^1\times \left((r_0-\epsilon,r_0) \cup\ (r_0,r_0+\epsilon)\right)\times \left(S^2\setminus \{p_0,p_\pi\}\right).
\end{equation}
The resulting manifold is diffeomorphic to 
\begin{equation}
\label{past/future}
S^3\times \left((r_0-\epsilon,r_0) \cup\ (r_0,r_0+\epsilon)\right).
\end{equation}
The fibration defined by the orbits of $\xi$ provides the projection
\begin{equation}
S^3\times \left((r_0-\epsilon,r_0) \cup\ (r_0,r_0+\epsilon)\right)\ \rightarrow\ S^2\times \left((r_0-\epsilon,r_0) \cup\ (r_0,r_0+\epsilon)\right)
\end{equation}
that can be factorised to the Hopf fibration 
\begin{equation}
S^3\ \rightarrow\ S^2.
\end{equation}

Finally, in order to extend (\ref{past/future}) across the horizon to make it 
\begin{equation}\label{complete}
S^3\times \left(r_0-\epsilon, r_0+\epsilon\right),
\end{equation}
we need to replace the coordinates $(t,r,\theta,\varphi)$ by the horizon penetrating coordinates $(v,r,\theta,\tilde{\phi})$
introduced in Sec. \ref{sec:horizonKNUT}.

\section{Relation with general Type D horizons and accelerated BHs}
Geometry of the Killing horizons can be described and studied in a manner independent of the bulk of surrounding spacetime.
The vacuum Einstein equations with a cosmological constant imply constraints.
If we add an assumption that the spacetime Weyl tensor is of the Petrov type D at the horizon and then we obtain an equation on the geometry of the space of null generators, the rotation connection $1$-form and the surface gravity.
A general solution to that equation was found in the axially symmetric case \cite{localnohairPhysRevD.98.024008,hopf}.
For the horizons of the Hopf fibration structure, there was derived a family of general solutions parametrised by $4$ parameters including the cosmological constant $\Lambda$.
The goal of this section is to identify among the general type D horizons, those that are embeddable in the Kerr-NUT-de Sitter spacetimes and set the $3$-parameter family of the geometries derived in Sec. \ref{sec:horizonKNUT}.
Next, in order to find the remaining general Type D horizons of the Hopf structure, we will have to extend our investigation to a more general class of spacetimes, namely so called accelerated Kerr–NUT–(anti-)de Sitter spacetimes. 

\subsection{Comparison with the abstractly derived Hopf bundle horizons}
\label{sec:comparison}
The theory of the abstract, unembedded type D horizons such that the null generators have the Hopf bundle structure provides the following family of geometries of the moduli space endowed with coordinates $x\in[-1,1]$ and $\phi\in[0,2\pi)$: 
\begin{equation}
\label{eq:metric-axial}
    q_{\rm gen}= R^2\bigg(\frac{1}{\tilde{P}_{\rm gen}(x)^2} dx^2 + \tilde{P}_{\rm gen}(x)^2 d\phi^2\bigg),
\end{equation}
where  $R$ is a constant (the area radius) and the function $\tilde{P}_{\rm gen}$ reads \cite{hopf}
\begin{equation}
    \tilde{P}_{\rm gen}^2=\frac{(1-x^2)\Big(\big(x-\tfrac{1}{2}\eta n(1-\tfrac{1}{6}\Lambda \gamma)\big)^2+\eta^2+\frac{1-x^2}{1-\tfrac{1}{6}\Lambda\gamma }\Big)}{\big(x-\tfrac{1}{2}\eta n(1-\tfrac{1}{6}\Lambda \gamma)\big)^2+\eta^2},
\end{equation}
where the parameter $\gamma$ is related to $R$, namely
\begin{equation}
R= \frac{\gamma}{\frac{\Lambda}{3}\gamma-2},     
\end{equation}
and $(\eta, \,n)$ are real parameters.
Clearly, the family of the geometries can be parametrised by $4$ free parameters $(R,\eta,n,\Lambda)$ modulo some inequalities ensuring that $P>0$ and the rotation-connection $1$-form is not singular. 
On the other hand, in Sec. \ref{sec:horizonKNUT} above, we derived the $3$ dimensional family of the geometries of Killing horizons in Kerr-NUT-de Sitter spacetimes described by the metric tensors (\ref{eq:metric-horizon}) (after rescaling $x^3$ to $\varphi$), namely:
\begin{equation}
\label{eq:metric-horizon-nonsingular}
     {}^{(2)}q_H=\Sigma_0 P^{-1}d\theta^2+ \Sigma_0^{-1}P\sin^2\theta P(0)^{-2}\rho^2_0 d\varphi^2.
\end{equation}
All those non-singular Killing horizons of Kerr-NUT-de Sitter spacetime are parametrised by triples $(a,l,r_0)$, while the cosmological constant $\Lambda$ is determined by the non-differentiability removability condition (\ref{eq:constraint-params}) and the mass parameter $m$ is given by (\ref{eq:mass-constraint}).
Our aim in this subsection is a comparison between the two families.
Obviously the $3$ dimensional family of the horizons embedded in the Kerr-NUT-de Sitter spacetimes is contained in the $4$ dimensional family of all the type D horizon structures with a non-trivial fibration by the null generators.
The question is what specifically distinguishes the embedded horizons from the more general ones (probably also embeddable, however in more general spacetimes - see the next subsection). 

What we have to do, is to recast a metric (\ref{eq:metric-horizon-nonsingular}) in the form (\ref{eq:metric-axial}).
The area 2-form of the metric (\ref{eq:metric-horizon-nonsingular}) is
\begin{equation}
\label{eq:area-2-form}
    {}^{(2)}\epsilon_H=\frac{\rho_0}{P(0)}\sin\theta d\theta\wedge d\varphi.
\end{equation}
Integrating the above we can compare the areas and get
\begin{equation}
    R^2=\frac{\rho_0}{P(0)}.
\end{equation}
Comparing the coefficients of $d\phi$ in (\ref{eq:metric-axial}) and $d\varphi$ in (\ref{eq:metric-horizon-nonsingular}) we have
\begin{equation}    
    \tilde{P}^2_H=\frac{\sin^2\theta P R^2}{\Sigma},
\end{equation}
where the positivity of $\tilde{P}^2_H$ is guaranteed by the positivity $P>0$. 
The last step is to solve for $x=x(\theta)$.
The implied equation
\begin{equation}
    \frac{R^2}{\tilde{P}_H^2}dx^2=\frac{\Sigma}{P} d\theta^2 \implies dx^2=\sin^2\theta d\theta^2
\end{equation}
with the boundary condition $x(0)=-1$ has a unique solution
\begin{equation}
    x=-\cos\theta.
\end{equation}

Comparing the coefficients of the powers of $x$ in $\tilde{P}_{\rm gen}$ and in $\tilde{P}_H$ we get
\begin{equation}
    \begin{split}
        \gamma&=-\frac{6}{\Lambda}=-2(a^2+2l^2+2r_0^2),\\
        n\eta&=\frac{l}{a},\\
       \eta&=\pm\frac{r_0}{a}.
    \end{split}
\end{equation}
In particular, it follows that 
\begin{equation}
\label{eq:RLambda}
R^2=\frac{3}{2\Lambda}. 
\end{equation}

The conclusion is, that a general vacuum, type D horizon of the Hopf fibration structure is embeddable in a Kerr-NUT-dS spacetime if and only if 
the cosmological constant $\Lambda$ and the area radius are related by (\ref{eq:RLambda}).
The question still persists concerning an embedding of the type D Hopf horizons characterised by arbitrary $\Lambda$ and $R$.
As we will see below, a $4$ dimensional family of them (perhaps all) are contained in generalised Plebański-Demiański black hole, where we gain an additional degree of freedom due to the acceleration $\alpha$.

\subsection{Generalised black hole}
\label{sec:gen-bh}
In this subsection we sketch a generalisation of our quest for conical singularity free black hole solutions with Hopf like topology, to spacetimes with an additional parameter $\alpha$ interpreted as the acceleration of the black hole along the axial symmetry axis. 
Contrary to the NUT solutions there is no known way of representing accelerating Schwarzschild black hole metric without conical singularity in at least one pole \cite{griffiths_podolsky_2009}. 
In the literature the accelerating Kerr–NUT–(anti-)de Sitter spacetime is also known as the generalised Plebański-Demiański black hole. 
Its metric can be expressed in a form similar to Kerr–NUT–(anti-)de Sitter with different constituting functions $P$ and $\mathcal{Q}$, and an additional conformal factor function $F$.
The metric can be expressed as follows \cite{griffiths_podolsky_2009}
\begin{equation}
\begin{split}
\label{eq:MetricGenBH}
    ds^2=\frac{1}{F^2}\bigg\{-\frac{\mathcal{Q}}{\Sigma}(dt-A d\phi)^2  +\frac{\Sigma}{\mathcal{Q}}dr^2 
+\frac{\Sigma}{P}d\theta^2+\frac{P}{\Sigma}\sin^2\theta(adt-\rho d\phi)^2\bigg\},
\end{split}
\end{equation}
where
\begin{equation}
    \begin{split}
        F&=1-\frac{\alpha}{\omega}(l+a\cos\theta)r,\\
        \Sigma&=r^2+(l+a\cos\theta)^2,\\
        A&=a\sin^2\theta+4l\sin^2\tfrac{1}{2}\theta,\\
        \rho&=r^2+(l+a)^2=\Sigma-aA,\\
        \mathcal{Q}&=\omega ^2 k-2mr +\epsilon r^2 - 2\frac{\alpha n}{\omega}r^3-(\alpha k + \frac{1}{3}\Lambda)r^4,\\
        P&=1-a_3\cos\theta-a_4\cos^2\theta,\\
        a_3 &= 2 \frac{\alpha a m} {\omega} - \alpha^2 a l k -\frac{4}{3} \Lambda a l ,\\
        a_4 &= -\alpha^2 a^2 k - \frac{1}{3}\Lambda a^2,\\
        \epsilon&= \frac{\omega^2 k}{(a^2 - l^2)} + 4 \frac{\alpha l m }{\omega}-(a^2+3l^2)(\alpha^2 k + \Lambda/3),\\
        n& = \frac{\omega^2 k l}{(a^2 - l^2)} - \frac{\alpha m (a^2 - l^2)}{\omega} + (a^2 - l^2) (\alpha^2 k + \Lambda/3),\\
        k&=\frac{1 + 2 \alpha l m/\omega - l^2 \Lambda}{3\alpha^2l^2 + \omega^2/(a^2-l^2)}.\\
    \end{split}
\end{equation}
Apart from parameters $\alpha$, $a$, $l$, $m$ and $\Lambda$, which have their typical interpretation, there is also introduced an additional free parameter $\omega$.
It can be set to any (non-zero) value provided that $a$ and $l$ are not both equal to 0, otherwise necessarily $\omega=0$.

\subsection{Metric of the space of orbits of Killing vector field}
For the generalised black hole we employ the same procedure as for the Kerr–NUT–(anti-)de Sitter and arrive at the metric on the space of orbit differing from (\ref{eq:metric-orbits-KNDS}) only by the conformal factor and the form of the functions $P$ and $\mathcal{Q}$. 
Because of that many general conclusions still hold.
In particular the generalisation to the Hopf bundle, analysis of the signature and the non-extremal limit are analogous to the Kerr–NUT–anti-de Sitter case.
The moduli metric is expressed as
\begin{equation}
\label{eq:metric-2-sec-gen-BH}
  q=\frac{1}{F^2}\bigg\{\frac{\Sigma}{\mathcal{Q}}(dx^1)^2+\frac{\Sigma}{P}(dx^2)^2+\frac{P \mathcal{Q}\sin^2(x^2) \rho_0^2\Sigma}{\mathcal{Q}\Sigma_0^2- P \sin^2(x^2) a^2(r^2-r_0^2)^2} (dx^3)^2\bigg\}.
\end{equation}
Calculation of the limit of ratio of radius and circumference is especially easy as the conformal factor cancels out. 
In the consequence we arrive again at the condition (\ref{eq:condition-for-diff}) that has to be satisfied by our new function $P$ defined in the previous section. 
The result, a generalisation of the condition (\ref{eq:constraint-params}) reads:
\begin{equation}
\label{eq:constraint-params-genBH}
\begin{split}
    r_0^2=\Big(3 \alpha ^2 a^4 l (\omega -\alpha  l m)+a^2 \left(6 \alpha ^3 l^4 m-6 \alpha ^2 l^3
   \omega +\Lambda  l \omega ^3-3 \alpha  m \omega ^2\right)+\\
   +l \left(-3 \alpha ^3 l^5 m+3
   \alpha ^2 l^4 \omega +2 \Lambda  l^2 \omega ^3-3 \alpha  l m \omega ^2-3 \omega
   ^3\right)\Big)\\
   \big(-3 a^2 \alpha ^2 l (\alpha  l m+2 \omega )+3 \alpha ^3 l^4 m+6 \alpha ^2
   l^3 \omega -2 \Lambda  l \omega ^3+3 \alpha  m \omega ^2\big)^{-1}.
   \end{split}
\end{equation}
Similarly it requires $a\neq 0$ and $l\neq 0$, but $\Lambda\neq0$ is relaxed. 
The condition (\ref{eq:constraint-params-genBH}) reduces to (\ref{eq:constraint-params}) for $\alpha=0$, regardless of chosen $\omega$, but this limit requires $\Lambda \neq 0$. 
We can simultaneously and unambiguously solve (\ref{eq:constraint-params-genBH}) and $\mathcal{Q}(r_0)=0$ for $\Lambda$ and $m$ to get the 4-dimensional family of horizons parametrized by $(a,l,r_0,\alpha)$:

\begin{equation}
\begin{split}
   m= -\big(&a^4 l \omega ^2-2 a^2 l^3 \omega ^2-2 a^2 \alpha  l^2 r_0^3 \omega -2 a^2 \alpha l r_0^3 \omega +2 a^2 l r_0^2 \omega ^2+l^5 \omega ^2+2 \alpha  l^4 r_0^3 \omega +2 \alpha  l^3 r_0^3 \omega -6 l^3 r_0^2 \omega ^2+\\
   &-4 \alpha  l^2 r_0^5 \omega +l r_0^4 \omega ^2\big)\big(-a^4 \alpha  l^2 \omega +2 a^4 \alpha ^2 l r_0^3-2 a^4 \alpha ^2
   r_0^3-a^4 \alpha  r_0^2 \omega +2 a^2 \alpha  l^4 \omega +4 a^2 \alpha ^2 l^3
   r_0^3+4 a^2 \alpha ^2 l r_0^5+\\
   &-2 a^2 l r_0 \omega ^2-2 a^2 \alpha ^2 r_0^5-2
   a^2 \alpha  r_0^4 \omega -\alpha  l^6 \omega -6 \alpha ^2 l^5 r_0^3+2 \alpha ^2 l^4
   r_0^3+5 \alpha  l^4 r_0^2 \omega -6 \alpha ^2 l^3 r_0^5-4 l^3 r_0 \omega ^2+\\
   &+2
   \alpha ^2 l^2 r_0^5+5 \alpha  l^2 r_0^4 \omega -4 l r_0^3 \omega ^2-\alpha 
   r_0^6 \omega \big)^{-1},
\end{split}
\end{equation}
\begin{equation}
    \begin{split}
    \Lambda=\big(&3 (2 a^6 \alpha ^4 l r_0^3-2 a^6 \alpha ^4 r_0^3-a^6 \alpha ^3 r_0^2 \omega +a^4 \alpha \omega ^3-6 a^4 \alpha ^4 l^3 r_0^3+6 a^4 \alpha ^4 l^2 r_0^3+3 a^4 \alpha ^3 l^2 r_0^2 \omega +4 a^4 \alpha ^4 l r_0^5+\\
   &-2 a^4 \alpha ^2 l r_0 \omega ^2-2 a^4 \alpha^4 r_0^5-2 a^4 \alpha ^3 r_0^4 \omega +6 a^2 \alpha ^4 l^5 r_0^3-6 a^2 \alpha^4 l^4 r_0^3-3 a^2 \alpha ^3 l^4 r_0^2 \omega -8 a^2 \alpha ^4 l^3 r_0^5+\\
   &+4 a^2\alpha ^2 l^3 r_0 \omega ^2-2 a^2 \alpha  l^2 \omega ^3+4 a^2 \alpha ^4 l^2 r_0^5+8 a^2 \alpha ^3 l^2 r_0^4 \omega -8 a^2 \alpha ^2 l r_0^3 \omega ^2-a^2 \alpha ^3 r_0^6 \omega +2 a^2 \alpha  r_0^2 \omega ^3+\\
   &-2 \alpha ^4 l^7 r_0^3+2 \alpha ^4 l^6 r_0^3+\alpha ^3 l^6 r_0^2 \omega +4 \alpha ^4 l^5 r_0^5-2 \alpha ^2 l^5 r_0 \omega ^2+\alpha  l^4 \omega ^3-2 \alpha ^4 l^4 r_0^5-6 \alpha ^3 l^4 r_0^4 \omega +8\alpha ^2 l^3 r_0^3 \omega ^2+\\
   &+\alpha ^3 l^2 r_0^6 \omega -6 \alpha  l^2 r_0^2 \omega
   ^3-2 \alpha ^2 l r_0^5 \omega ^2+2 l r_0 \omega ^4+\alpha  r_0^4 \omega
   ^3)\big)\big
   (\omega ^2 (a^4 \alpha  l^2 \omega -2 a^4 \alpha ^2 l r_0^3+2 a^4 \alpha ^2
   r_0^3+a^4 \alpha  r_0^2 \omega +\\
   &-2 a^2 \alpha  l^4 \omega -4 a^2 \alpha ^2 l^3
   r_0^3-4 a^2 \alpha ^2 l r_0^5+2 a^2 l r_0 \omega ^2+2 a^2 \alpha ^2 r_0^5+2
   a^2 \alpha  r_0^4 \omega +\alpha  l^6 \omega +6 \alpha ^2 l^5 r_0^3-2 \alpha ^2 l^4 r_0^3+\\
   &-5 \alpha  l^4 r_0^2 \omega +6 \alpha ^2 l^3 r_0^5+4 l^3 r_0 \omega ^2-2
   \alpha ^2 l^2 r_0^5-5 \alpha  l^2 r_0^4 \omega +4 l r_0^3 \omega ^2+\alpha 
   r_0^6 \omega )\big)^{-1},
    \end{split}
\end{equation}
both of which reduce to (\ref{eq:mass-constraint}) and (\ref{eq:constraint-Lambda}) for $\alpha=0$, independently of chosen $\omega$.
Then using parametrization $(a,l,r_0,\alpha)$ the rescaling constant is
\begin{equation}
\begin{split}
    c=\frac{1}{P(0)}=&\Big(\alpha  a^4 \big(l^2 \omega -2 \alpha  (l-1) r_0^3+r_0^2 \omega \big)+2 a^2
   \big(-\alpha  l^4 \omega -2 \alpha ^2 l^3 r_0^3+l (r_0 \omega ^2-2 \alpha ^2
   r_0^5)+\alpha  r_0^4 (\alpha  r_0+\omega
   )\big)+\\
   &+(l^2+r_0^2) (\alpha  l^4 \omega +6 \alpha ^2 l^3 r_0^3-2 \alpha 
   l^2 r_0^2 (\alpha  r_0+3 \omega )+4 l r_0 \omega ^2+\alpha  r_0^4 \omega
   )\Big)\\
   &\Big(\big(a^2+2 a l+l^2+r_0^2\big) \big(\alpha  a^4 \omega +2 \alpha  a^2 (-l^2
   \omega +r_0^3 (\alpha -3 \alpha  l)+r_0^2 \omega )+\alpha  l^4 \omega +6 \alpha ^2
   l^3 r_0^3+\\
   &-2 \alpha  l^2 r_0^2 (\alpha  r_0+3 \omega )+4 l r_0 \omega^2+\alpha  r_0^4 \omega \big)\Big)^{-1}.
\end{split}
\end{equation}

It should be stressed that again the rescaling constant depends only on the value of $r_0$ of the horizon and not on the coordinate $r$.
This property makes it possible to remove the singularity at the horizon and in its future/past simultaneously.

It should be noted, though, that introduction of $\alpha$ does not lead to qualitatively new solutions to the non-differentiability removability condition:
there is no solution with $a=0$ and $\alpha\neq0$ thus we recover the fact there is no known solution of accelerating NUT spacetime \cite{griffiths_podolsky_2009}.
Also there is no non-singular accelerating Kerr–NUT–(anti-)de Sitter: if we put $l=0$ then $\omega=a$ and (\ref{eq:condition-for-diff}) requires that $\alpha m = 0$.

\section{Summary}
We have studied the spaces of orbits of horizon forming Killing vector fields in the Kerr–NUT–(anti-)de Sitter spacetimes (\ref{eq:MetricKNdS}).
For every Kerr-NUT-(A)dS spacetime and every Killing horizon, the space of the null generators is a (topological) $2$ dimensional sphere $S^2$ endowed with an induced axisymmetric metric tensor ${}^{(2)}q_H$.
The metric tensor ${}^{(2)}q_H$ has a singularity at one or both the poles, for generic values of the parameters $(m,a,l,\Lambda)$ defining a Kerr-NUT-(A)dS spacetime.
We have showed, that the singularity is removable (simultaneously at the two poles) if and only if 
\begin{equation}  
\label{Lambda}
        \frac{3}{\Lambda}=a^2+2l^2+2r_0^2,
    \end{equation}
where $r_0$ is a root of the function $\mathcal{Q}$ (\ref{eq:functions-metric}) that defines the horizon.
The non-singular configurations can be freely parametrised by $(r_0,a,l)\in \mathbb{R}^3$, while the cosmological constant $\Lambda$ and the mass parameter $m$ are determined by  (\ref{Lambda}) and (\ref{eq:mass-constraint}).
The area radius of every non-singular horizon in a Kerr-NUT-dS spacetime is
\begin{equation}
R = \sqrt{\frac{3}{2\Lambda}}. 
\end{equation}       
The orbit space of a Killing vector field in a neighbourhood of a Killing horizon in Kerr-NUT-(A)dS spacetime is topologically $S^2\times \mathbb{R}$ endowed with an induced axisymmetric metric tensor $q$ whose nature critically depends on the multiplicity of the root $r_0$ and on a side of the horizon.
In the current work we considered the single multiplicity case, that is a non-extremal horizon.
Generically, the metric tensor $q$ has a conical singularity at the rotational symmetry axis (in addition to the change of the signature while passing the horizon).
Removing the conical singularity at the Killing horizon (whenever possible) removes it also in all the neighbourhood.
Hence, the non-singular Killing horizons in the Kerr-NUT-dS spacetimes are surrounded by non-singular neighbourhoods.
The topology of suitably extended neighbourhoods is $S^3\times \mathbb{R}$ and the topology of the horizons themselves is $S^3$.
By modifying the global glueing conditions, $S^3$ could be replaced by any $U(1)$ bundle ovever $S^2$.
Given a non-singular Killing horizon in a Kerr-NUT-dS spacetime, the catch is, that every other Killing horizon in that spacetime suffers irremovable conical singularity, since the condition (\ref{Lambda}) can be satisfied for at most one Killing horizon at once.

\textcolor{black}{A remark on the uniqueness of the extension of the spacetime around the horizon is in order. Once a horizon in a the (possibly accelerated) Kerr-NUT-(a)dS spacetime is extended in the non-singular way, the existence and uniqueness problem of a suitable extension of the neighbourhood could be approached by using a combination of recent results by Rácz and collaborators, namely the black hole holograph construction \cite{Racz_2007} and the characteristic initial value problem for Killing spinors \cite{Cole_2018}.
Given a sufficiently small segment of the horizon, the black hole holograph creates locally a vacuum spacetime with the same cosmological constant as the original, that contains the segment.
The Killing spinor characteristic data theorem implies that the hologram spacetime is of the Petrov type D. However glueing those spacetimes consistently into manifold would require some work, and still the result would be implicit, rather than explicit as the solutions studied in the current paper.}

We have also generalised those results to the accelerated Kerr–NUT–(anti-)de Sitter spacetimes parametrised freely by $5$ parameters, namely the previous $m,a,l,\Lambda$ and a new parameter $\alpha$ (acceleration) (\ref{eq:MetricGenBH}).
The Killing horizon singularity removability condition takes the form (\ref{eq:constraint-params-genBH}).
The family of the parameters $(m,a,l,\Lambda,\alpha)$ that satisfy that condition is $4$ dimensional and parametrised by $(r_0, a, l,\alpha$).
                    
The current results on non-singular Killing horizons in the 
Kerr-NUT-(A)dS spacetimes are compared with the results of the local theory of Killing horizon like null surfaces \cite{hopf}.
The local theory provides a $4$ dimensional family of axisymmetric vacuum the Petrov type D horizon geometries of the topology of the Hopf fibration $S^3\rightarrow S^2$.
We identified among them the 3 dimensional family of the horizons embeddable in a Kerr-NUT-dS spacetime by an explicit calculation.
They are distinguished by the relation (\ref{eq:RLambda}) between the area radius and $\Lambda$.
The accelerated Kerr-NUT-(A)dS spacetimes contain a $4$-parameter family of the type D horizons hence the numbers of the parameters agree, however an explicit $1$-$1$ correspondence between the members of the two families is an open problem. 

\textit{Acknowledgements} Special thanks are due to István Rácz who pointed out to us that every vacuum type D horizon, including  topologically that of Hopf, should be
embeddable in one of the type D vacuum spacetimes. The authors also wish to thank Remigiusz Durka for inspirational conversations and Marc Mars for consulting our results and pointing out similar work done in the field. This work was partially supported by the Polish
National Science Centre grants No. 2017/27/B/ST2/02806 and No.
2016/23/P/ST1/04195.

\appendix
\section{Explicit differentiation of the resulting non-singular metrics}
\label{sec:appendix}
In this section we show by inspection that the horizon metric (\ref{eq:metric-horizon-nonsingular}) and the metric (\ref{eq:metric-neighbourhood-nonsingular}) on the horizon's future/past induced on the surface $x^1=r$ are smooth. Here we use the assumption that the coordinates $(\theta,\varphi)$ are of the spherical nature, defined on the $[0,\pi] \times [0,2\pi)$. Subsequently we treat the metrics as the metrics defined on the unit spheres and project them onto the equatorial plane. This corresponds to using the spherical coordinates:
\begin{equation}
    \begin{split}
        x&=\sin\theta\cos\varphi,\\
        y&=\sin\theta\sin\varphi,\\
        z&=\cos\theta.
    \end{split}
\end{equation}
We have to restrict this projection by choosing only one half of the sphere, covering only one of the suspicious poles at a time. Then 
\begin{equation}
    \begin{split}
        \sin\theta&=\sqrt{x^2+y^2},\\
        \cos\theta&=\epsilon\sqrt{1-x^2-y^2},\\
        d\theta&=\epsilon\frac{xdx+ydy}{\sqrt{1-x^2-y^2}\sqrt{x^2+y^2}},\\
        d\varphi&=\frac{xdy-ydx}{x^2+y^2},
    \end{split}
\end{equation}
where we use $\epsilon=1$ when projecting the northern hemisphere, i.e. when $\theta\in[0,\pi/2]$ and $\epsilon=-1$ for the southern, i.e. when $\theta\in[\pi/2,\pi]$.
Using the above projection we obtain the following metrics
\begin{equation}
\begin{split}
&{}^{(2)}q_H=\left(\left(a^2+2 \left(l^2+r_0^2\right)\right) \left((x dy-y dx)^2 \left(a^2
   \left(2-x^2-y^2\right)+4 a l \epsilon\sqrt{1-x^2-y^2}+2 \left(l^2+r_0^2\right)\right)^2 \right.\right.+ \\
   &\left. \left. +4 (x dx
   +y dy)^2 \left(\left(a \epsilon\sqrt{1-x^2-y^2}+l\right)^2+r_0^2\right)^2\right)\right)
   \Bigg(4 \left(x^2+y^2\right) \left(8 a^3 l \epsilon\sqrt{1-x^2-y^2} \left(2-x^2-y^2\right)+2
   \left(l^2+r_0^2\right)\right)\\
   &\left(\left(a \epsilon\sqrt{1-x^2-y^2}+l\right)^2+r_0^2\right)\Bigg)^{-1}
\end{split}
\end{equation}

\begin{equation}
    \begin{split}
        &{}^{(2)}q=-\left[\left(a^2+2 \left(l^2+r_0^2\right)\right) \left(\left(a
   \epsilon\sqrt{1-x^2-y^2}+l\right)^2+r^2\right) \right.\\
   &\left.\left((r_0-r) \left(a^4-2 a^2\left(l^2-r_0^2\right)+l^4+2 l^2 r_0 (2 r-r_0)+r r_0 \left(r^2+r
   r_0-r_0^2\right)\right) (dy x-dx y)^2 \left(a^2 \left(-\left(2-x^2-y^2\right)\right)+\right.\right.\right.\\&
   \left.\left.\left.+4 a l \epsilon\sqrt{1-x^2-y^2}+2
   \left(l^2+r_0^2\right)\right)^2\right)\left(a^2 r_0 \left(r^2-r_0^2\right)^2
   \left(x^2+y^2\right) \left(a^2 \left(2-x^2-y^2\right)-4 a l \epsilon\sqrt{1-x^2-y^2}+\right.\right.\right.\\&\left.\left.\left.-2
   \left(l^2+r_0^2\right)+
   -(r-r_0) \left(a^4-2 a^2 \left(l^2-r_0^2\right)+l^4+2
   l^2 r_0 (2 r-r_0)+r r_0 \left(r^2+r r_0-r_0^2\right)\right)
   \right.\right.\right.\\ &\left.\left.\left.
   \left(a^2 \left(-\left(1-x^2-y^2\right)\right)+2 a l \epsilon\sqrt{1-x^2-y^2}+l^2+r_0^2\right)^2\right)^{-1}+
   \right.\right.\\&\left.\left.
   -\frac{4(dx x+dy y)^2}{1-x^2-y^2}\right)\right] \left[4\left(x^2+y^2\right) \left(a^2
   \left(2-x^2-y^2\right)-4 a l \epsilon\sqrt{1-x^2-y^2}-2 \left(l^2+r_0^2\right)\right)\right]^{-1}
    \end{split}
\end{equation}
One can verify that the above metrics gives rise to the components $q_{xx},q_{xy},q_{yy}$ that are constituted only of the sums and products of functions of $x$ and $y$ that are smooth at the point $(x,y)=(0,0)$.
Thus we conclude that the metrics ${}^{(2)}q_H$ and ${}^{(2)}q$ are smooth at the poles $\theta=0$ and $\theta=\pi$ corresponding to the point $(x,y)=(0,0)$ when projecting onto the suitable hemisphere.
Incidentally the metric in not even continuous at the poles without introducing the rescaling $1/ P(0)$.

In the similar fashion one can check explicitly that the rotation-connection 1-forms $d\tau+\omega(r,x^3) d\varphi$ defined by (\ref{eq:rot-1-form}) as well as the 1-form after the coordinate change (\ref{eq:trans}) are smooth, when projected onto the suitable hemisphere.
The problematic part of the 1-form is 
\begin{equation}
\begin{split}
    &\omega(r,\theta)d\phi=\bigg(3 \left(a^2+2 \left(l^2+r_0^2\right)\right) (dy x-dx y) \left(\left(a
   \left(x^2+y^2\right)+l \left(\sqrt{-x^2-y^2+1}-1\right)^2\right)\right. \\
   &\left.\left(-\frac{1}{3} \Lambda  \left(a^2 \left(3 l^2+r^2\right)-3 l^4+6 l^2 r^2+r^4\right)+a^2-l^2-2 m r+r^2\right) \left(-a \left(a
   \left(x^2+y^2\right)+l \left(\sqrt{-x^2-y^2+1}-1\right)^2\right)+\right.\right.\\
   &\left.\left.+(a+l)^2+r_0^2\right)+\frac{1}{3} a \left(r^2-r_0^2\right) \left(x^2+y^2\right) \left((a+l)^2+r^2\right) \left(a^2 \Lambda  \left(-\left(x^2+y^2-1\right)\right)+4 a \Lambda  l \sqrt{-x^2-y^2+1}+3\right)\right)\bigg)\\
   &\bigg(2 \left(x^2+y^2\right) \left(\left(a^2 \left(3 \Lambda  l^2+\Lambda  r^2-3\right)-3 \Lambda  l^4+l^2 \left(6 \Lambda  r^2+3\right)+r \left(6 m+\Lambda  r^3-3 r\right)\right)\right.\\
   &\left.\left(a^2 \left(-\left(x^2+y^2-1\right)\right)+a l \left(2 \sqrt{-x^2-y^2+1}+x^2+y^2\right)+l^2+r_0^2\right)^2+\right.\\
   &\left.-a^2 \left(r^2-r_0^2\right)^2
   \left(x^2+y^2\right) \left(a^2 \Lambda  \left(x^2+y^2-1\right)-4 a \Lambda  l
   \sqrt{-x^2-y^2+1}-3\right)\right)\bigg)^{-1}
\end{split}
\end{equation}
The above 1-form is also composed of the function smooth at the $(x,y)=(0,0)$ except for the part proportional to 
\begin{equation*}
    \frac{\left(1-\sqrt{1-x^2-y^2}\right)^2}{x^2+y^2}=2\frac{1-\sqrt{1-x^2-y^2}}{x^2+y^2}-1=:2 f(x,y)-1.
\end{equation*}
The limit of the function $f$ to the pole exists and evaluates to
\begin{equation*}
    \lim _{(x,y)\to(0,0)} f(x,y)=\tfrac{1}{4}.
\end{equation*}
Differentiating $f$ we get
\begin{equation}
    \begin{split}
        &f_{,x}=-\frac{x \left(2 \sqrt{1-x^2-y^2}+x^2+y^2-2\right)}{\sqrt{1-x^2-y^2} \left(x^2+y^2\right)^2}=\frac{f^2 x}{\sqrt{1-x^2-y^2}},\\
        &f_{,y}=-\frac{y \left(2 \sqrt{1-x^2-y^2}+x^2+y^2-2\right)}{\sqrt{1-x^2-y^2} \left(x^2+y^2\right)^2}=\frac{f^2 y}{\sqrt{1-x^2-y^2}}.      
    \end{split}
\end{equation}
And so the the derivative of $f$ at zero exists and is composed only of the smooth functions and the function $f$ which is at least once differentiable at zero.
We conclude that all derivatives of $f$ at $(x,y)=(0,0)$ exist and $f$ is smooth, making the whole rotation-connection 1-form smooth at the poles.
Similar analisys can be made for the 1-form $\omega'(r',\theta')$ projected onto the southern hemisphere.

\bibliographystyle{unsrt}
\bibliography{bibliography}

\end{document}